\documentclass[10 pt,twocolumn]{autart} %

\usepackage[USenglish]{babel}
\usepackage{amsmath} %
\usepackage{amssymb}  %
\usepackage{algpseudocode}      %
\usepackage{tikz,pgfplots}
\pgfplotsset{compat=newest}
\usepgfplotslibrary{external}           %
\usepackage[utf8x]{inputenc}
\usepackage{graphicx}

\tikzset{%
    font=\small
}
\pgfplotsset{
  title style={font=\small}
}

\newcommand{\R}{\mathbb{R}}
\newcommand{\PP}{\mathbb{P}}
\newcommand{\PE}{\mathbb{E}}
\newcommand{\X}{\mathbb{X}}

\newcommand{\U}{\mathbb{U}}
\newcommand{\V}{\mathbb{V}}
\newcommand{\W}{\mathbb{W}}
\newcommand{\D}{\mathbb{D}}
\newcommand{\C}{\mathbb{C}}
\newcommand{\N}{\mathbb{N}}
\newcommand{\AB}{\mathbb{G}}%
\renewcommand{\l}{\ell}

\begin{document}
\begin{frontmatter}

\title{%
Stochastic MPC with Offline Uncertainty Sampling\thanksref{footnoteinfo}%
}

\thanks[footnoteinfo]{%
M. Lorenzen and F. Allg\"ower would like to thank the German Research Foundation (DFG) for financial support of the project within the Cluster of Excellence in Simulation Technology (EXC 310/2) at the University of Stuttgart.\\
The research is partially funded by the CNR Joint International Lab COOPS.
}

\author[Stutt]{Matthias Lorenzen}\ead{matthias.lorenzen@ist.uni-stuttgart.de}, 
\author[Torino]{Fabrizio Dabbene}\ead{fabrizio.dabbene@ieiit.cnr.it}, 
\author[Torino]{Roberto Tempo}\ead{roberto.tempo@polito.it}, 
\author[Stutt]{Frank Allg\"ower}\ead{frank.allgower@ist.uni-stuttgart.de}%
\address[Stutt]{Institute for Systems Theory and Automatic Control, University of Stuttgart, %
Germany}%
\address[Torino]{CNR-IEIIT, Politecnico di Torino, %
Italy}%


\begin{abstract}
  For discrete-time linear systems subject to parametric uncertainty described by random variables, we develop a sampling-based Stochastic Model Predictive Control algorithm.
  Unlike earlier results employing a scenario approximation, we propose an offline sampling approach in the design phase instead of online scenario generation. 
  The paper highlights the structural difference between online and offline sampling and provides rigorous bounds on the number of samples needed to guarantee chance constraint satisfaction. %
  The approach does not only significantly speed up the online computation, but furthermore allows to suitably tighten the constraints to guarantee robust recursive feasibility when bounds on the uncertain variables are provided. Under mild assumptions, asymptotic stability of the origin can be established.%
\end{abstract}

\end{frontmatter}
\section{Introduction}
In recent years, Stochastic Model Predictive Control (SMPC) has received significant attention for constrained control of systems, where a probabilistic description of disturbances and uncertainties is given~\cite{Kouvaritakis2016_MPCBook}. %
The online solution of stochastic constrained control remains a challenging problem, which, among other methods, has been addressed using probabilistic approximations, for which significant progress has been made over the last two decades~\cite{Tempo2012_RandAlgForAnalysisAndDesign}.
  These probabilistic approximations often employ sampling-based methods, which have the advantage of being computationally tractable and are able to handle general types of stochastic uncertainty, including parametric uncertainty that enters nonlinearly in the system dynamics and constraints or has a non-convex support. 
  For stochastic optimization problems that are convex in the optimization variables, in~\cite{Calafiore2006_ScenarioApproachRobContrDesign} and subsequently~\cite{Calafiore2010_RandomConvexPrograms,Campi2008_ExactFeasibilityOfRandSolOfUncertainConvexProgs,Campi2011_SampleAndDiscardApprTpCCOpt} the authors studied the so-called ``scenario approach''. Tight bounds on the necessary number of samples were derived to provide probabilistic guarantees that the \emph{solution} of the approximate sampling-based program satisfies the original chance constrained problem~\cite{Alamo2009_RandStrategiesforProbSolOfUncertaintFeasAndOptProbs,Calafiore2010_RandomConvexPrograms,Campi2008_ExactFeasibilityOfRandSolOfUncertainConvexProgs,Campi2011_SampleAndDiscardApprTpCCOpt}.
  These results have been applied to SMPC for linear systems with stochastic uncertainty, e.g. \cite{Calafiore2013StochasticMPC}%
  , and are shown to be less conservative compared to robust outer approximations~\cite{Deori2014_CompApproachesToRMPC}.
  While sampling allows for nearly arbitrary uncertainty in the system, two key issues %
  are (i) the large number of constraints that are generated, which increases the computational and memory requirements significantly~\cite{Zhang2014_OnSampleSizeOfRandMPCwAppl}, thereby reducing its applicability in SMPC~\cite{Mesbah2014_StochasticNonlinearMPC}, and (ii) the lack of guaranteed recursive feasibility~\cite{Deori2014_CompApproachesToRMPC,Schildbach2014_ScenarioApproachForSMPC}.

  The first problem has been addressed through exploiting structural properties of the optimization program in SMPC, e.g.~\cite{Schildbach2014_ScenarioApproachForSMPC,Zhang2014_OnSampleSizeOfRandMPCwAppl}. In~\cite{Schildbach2014_ScenarioApproachForSMPC} only the first step violation probability is considered, which provides bounds on the closed-loop constraint violation, but increases the probability of not being recursively feasible. Furthermore, the lack of rigorous guarantees for asymptotic stability is a critical point.
  Structural problems of recursive feasibility in SMPC have been highlighted in~\cite{Primbs2009_StochRecedingHorizonContrOfConstrLinSysWithMultNoise} and a solution using a combination of robust and stochastic constraint tightening for linear systems with parametric disturbances has been proposed in~\cite{Fleming2014_StochasticTubeMPCforLPVSysWithSetInclCond}.%

  The main contribution of this paper is to propose a solution that alleviates the two mentioned disadvantages. %
  The SMPC controller design is based on a sampling approach in the \emph{offline} design, as suggested in~\cite{Mayne2014_MPCSurvey}.
  In contrast to previously mentioned online sampling approaches, scenarios are generated offline and only \emph{necessary} samples are kept for online optimization. %
  Since the tight bounds derived in the scenario approach cannot be directly used, we reformulate the problem to exploit results from statistical learning theory. We provide precise statements on the necessary sample complexity such that, within a user-specified confidence, constraint satisfaction is guaranteed.
  Following an approach similar to~\cite{Lorenzen2015_improvedConstrTighteningForSMPC,Lorenzen2015_ConstrTighteningAndStabInSMPC}, we extend the idea of an additional first-step constraint to guarantee recursive feasibility.
  Conditions are stated under which the closed-loop system is asymptotically stable with probability one.

  Preliminary results have been presented in the conference paper~\cite{Lorenzen2015_ScenarioBasedStochasticMPC}, %
 where an offline sampling approach has been shown to be advantageous with respect to computational requirements and recursive feasibility, but lacking rigorous bounds on the sample complexity. In contrast, here we first illustrate the difference between online and offline uncertainty sampling in receding horizon control and based thereon provide bounds on the number of samples to guarantee chance constraint satisfaction when the uncertainty is sampled offline. 
 A first step constraint to guarantee recursive feasibility has previously been introduced in~\cite{Lorenzen2015_ConstrTighteningAndStabInSMPC}. The paper introduces a non-conservative constraint tightening and discusses the relation to performance and stability in Stochastic MPC for systems with additive disturbance. 

  The remainder of this paper is organized as follows. Section~\ref{sec:ProbSetup} introduces the scenario receding horizon problem and briefly reviews the underlying theory. In Section~\ref{sec:offlineSampling} the difference between offline and online sampling is highlighted and suitable bounds on the sample complexity are derived.
  Based thereon, the SMPC design is given, starting with a suitable cost and constraint reformulation, followed by the derivation of additional constraints for recursive feasibility. The section ends with the complete SMPC algorithm and closed-loop properties. %
  Finally, Section~\ref{sec:Concl} provides some conclusions and proposes directions for future work.

\paragraph*{Notation}
The notation employed is standard. Uppercase letters are used for matrices and lower case for vectors. $[A]_j$ and $[a]_j$ denote the $j$-th row and entry of the matrix $A$ and vector $a$, respectively. 
Blackboard boldface letters (e.g. $\W$) denote sets, with the exception of $\PP$ and $\PE$ for probability measure and expected value. $\PP^n$ denotes the $n$-fold product probability measure, and $\operatorname{supp}(\cdot)$ the support of a random variable.
The set of integers from $i$ to $j$ is denoted by $\N_i^{j} := \{i,i+1,\dots,j\}$.
We use the double index notation to denote predicted states and inputs, e.g. $x_{l|k}$ for state $x$ predicted $l$ steps ahead from time $k$ with $x_{0|k} = x_k$. Bold letters, e.g. $\mathbf x_k = [x_{0|k}^\top \dots x_{T-1|k}^\top]^\top$, are used to denote the stack vector of $T$ predicted values.

\section{Problem Setup and Preliminary Results}\label{sec:ProbSetup}
The uncertain discrete-time system to be controlled is given by
\begin{equation}
  x_{k+1} = A(q_k) x_k + B(q_k) u_k %
  \label{eqn:xUncertainSystem}
\end{equation}
with state $x_k \in \R^{n}$, control input $u_k \in \R^{m}$ and uncertainty $q_k \in \R^{n_q}$.
The system matrices $A(q_k)$ and $B(q_k)$, of appropriate dimensions, are subject to stochastic parametric disturbance. The parameter $q_k$ can enter nonlinearly under the following assumption on $A(q_k)$ and $B(q_k)$.
\begin{assum}[Stochastic Uncertainty]\label{ass:uncertaintySetSys}
The parameters $q_k \in \R^{n_q}$ for $k \in \N$ are realizations of independent and identically distributed (iid) multivariate, real valued random variables $Q_k$.
 Let $\AB = \{ (A(q_k),B(q_k))\}_{q_k \in \operatorname{supp}(Q_k)}$, a polytopic outer approximation $\bar \AB = \operatorname{co}\{(A^j,B^j)_{j\in \N_1^{N_c}}\}\supseteq \AB$ exists and is known.
\end{assum}
\begin{rem}
For the sake of robust recursive feasibility, we assume a known bound on the set $\AB$. This assumption could be relaxed to e.g. a confidence region for the parameters, which subsequently leads to a notion of probabilistic recursive feasibility. Since the robust outer bound is used only for a one-step prediction, choosing a large outer bound does not lead to as conservative constraint tightening as in Robust MPC.
\end{rem}

The system is subject to individual chance constraints %
on the state and hard constraints on the input
\begin{subequations}
  \begin{align}
    \PP\{ [H_x]_j  x_{l|k} &\le 1 ~|~ x_k \} \ge 1-\varepsilon_j, \quad &&\forall l\in \N_+ \label{seqn:probConstraints} \\
    H_u  u_{l|k} &\le 1, \quad &&\forall l\in \N \label{seqn:inputConstraints}
  \end{align}
  \label{eqn:origConstraints}%
\end{subequations}%
\noindent
with $\varepsilon_j \in (0,1)$ for $j \in \N_1^{p}$.

The control objective is to design a stabilizing receding horizon control, which guarantees constraint satisfaction and minimizes $J_\infty$, the expected value of an infinite horizon quadratic cost
\begin{equation}
  J_\infty = \sum_{i=0}^\infty \PE \left\{x_i^\top Qx_i + u_i^\top Ru_i \right\}
  \label{eqn:infHorizonCost}
\end{equation}
with $Q\in \R^{n\times n}$, $Q \succ 0$, $R\in \R^{m\times m}$, $R \succ 0$.

\subsection{Online Sampling-based SMPC algorithm}
The SMPC design we propose is closely related to the previously mentioned Scenario MPC~\cite{Calafiore2013StochasticMPC,Schildbach2014_ScenarioApproachForSMPC}, where the uncertainty is sampled \emph{online}. We briefly recall the most relevant results and assumptions of this approach.

The design relies on the standard assumption of the existence of a suitable terminal set $\X_T$ %
and an asymptotically stabilizing control gain for \eqref{eqn:xUncertainSystem}.
\begin{assum}[Terminal Set] \sloppy
  There exists a terminal set $\X_T \allowbreak=\allowbreak \{x ~|~ {H_T x \le 1}\}$, which is robustly forward invariant for~\eqref{eqn:xUncertainSystem} under the control law~$u_k=Kx_k$. Given any $x_k \in \X_T$, 
  the state and input constraints~\eqref{eqn:origConstraints} are satisfied and there exists $P \in \R^{n\times n}$  such that
\begin{equation*}
  Q + K^\top R K + \PE\left[ A_{cl}(Q_k)^\top P A_{cl}(Q_k) \right] - P \preceq 0
\end{equation*}
with $A_{cl}(Q_k) = A(Q_k) + B(Q_k)K$.
\end{assum}

A parameterized feedback policy
\begin{equation}
  u_{l|k} = Kx_{l|k} + v_{l|k}
  \label{eqn:prestabContr}
\end{equation}
is employed, where, for a given $x_{0|k} = x_k$, the correction terms $\{v_{l|k}\}_{l\in\N_{0}^{T-1}}$ are determined by the SMPC algorithm as the minimizer of the expected finite horizon cost %
\begin{multline}
    J_T(x_k,\mathbf{v}_{k}) \\
    = \PE\left\{ \sum_{l=0}^{T-1} \left( x_{l|k}^\top Qx_{l|k} + u_{l|k}^\top Ru_{l|k} \right) + x_{T|k}^\top P x_{T|k} ~|~ x_k\right\}.
  \label{eqn:finiteHorizonCostFnc}
\end{multline}
Note that $J_T(x_k,\mathbf{v}_k)$ is a convex function in $\mathbf v_k$ which can be computed explicitly, as shown in Section~\ref{sec:MainRes}.

Instead of evaluating the chance constraints~\eqref{seqn:probConstraints} directly, in~\cite{Calafiore2013StochasticMPC,Schildbach2014_ScenarioApproachForSMPC} a sampling-based approximation is used.
Let $\mathbf Q_k=\{Q_k,Q_{k+1},\dots, Q_{k+T-1}\}$ and let $\mathbf{q}^{(i)}_k=\{q^{(i)}_{0|k} ,q^{(i)}_{1|k},\dots, q^{(i)}_{T-1|k}\}$, $i\in\N_1^{N_s}$ denote randomly drawn samples from $\mathbf Q_k$. %
The \emph{finite horizon scenario program} is
\begin{subequations}
  \label{eqn:SampleBasedMPCOptimization}
  \begin{align}
    \min_{\mathbf v_k} ~& J_T\left( x_k,\mathbf v_k \right) \label{seqn:SampleBasedMPCCost}\\
    \text{ s.t. }
    & x_{l+1|k}^{(i)} = A(q^{(i)}_{l|k})x_{l|k}^{(i)} + B(q^{(i)}_{l|k})u_{l|k}^{(i)}, \quad x_{0|k}^{(i)}=x_k\nonumber\\
    & u_{l|k}^{(i)}=Kx_{l|k}^{(i)}+v_{l|k}\nonumber\\
    &\begin{aligned}
      & H_x x_{l|k}^{(i)} \le 1 \quad &&\forall l \in \N_1^{T-1}\\ 
      & H_u u_{l|k}^{(i)} \le 1 \quad &&\forall l \in \N_0^{T-1}\\
      & H_T x_{T|k}^{(i)} \le 1 \\
      & \text{for all } i \in \N_1^{N_s}.
    \end{aligned}\label{seqn:SampleBasedMPCConstr}
  \end{align} 
\end{subequations}

As shown in the following proposition, probabilistic guarantees that the solution to~\eqref{eqn:SampleBasedMPCOptimization} satisfies the original chance constraints can be derived, see~\cite{Alamo2015_RandAlgForDesignOfUncertainSys,Calafiore2010_RandomConvexPrograms,Campi2008_ExactFeasibilityOfRandSolOfUncertainConvexProgs,Tempo2012_RandAlgForAnalysisAndDesign} for the underlying theory, and~\cite{Alamo2015_RandAlgForDesignOfUncertainSys} for the particular sample complexity given below.
\begin{prop}[Scenario Program]\label{thm:scenario}
  Given $f(x,\theta):(\R^d \times S_\theta) \rightarrow \R$, convex in $x$ for any fixed $\theta \in S_\theta$ and assume $\theta \sim \PP$ with support $S_\theta$. Let $\Theta = \{\theta^{(1)}, \dots,\theta^{(N_s)}\}$ be a multisample of $\theta$ with the sample complexity
  \begin{equation}
    N_s \ge N(d,\varepsilon,\delta) = \frac{1}{\varepsilon}\frac{\e}{\e-1}\left( \ln\frac{1}{\delta} + (d-1) \right)
    \label{eqn:numberOfSamples}
  \end{equation}
  for $\varepsilon,\delta\in(0,1)$ and $\e$ being the Euler's number. For
  \begin{equation*}
    \begin{aligned}
      x^* = \arg \min_x &~c^\top x  \\
      \text{s.t. } & f(x,\theta^{(i)}) \le 0, \quad i\in\N_1^{N_s},
    \end{aligned}
  \end{equation*}
  with confidence $1-\delta$, it holds $\PP\{f(x^*,\theta)\le0\}\ge 1-\varepsilon$.
\end{prop}

This result can be applied to receding horizon control, in particular the finite horizon program~\eqref{eqn:SampleBasedMPCOptimization}, leading to the following corollary, originally derived in~\cite[Proposition 3.1 b)]{Calafiore2013StochasticMPC}.
\begin{cor}\label{prop:NumberOfSamples}
  Let $N_s\ge N(Tm,\varepsilon,\delta)$ with $\varepsilon \le \varepsilon_j$ be chosen according to~\eqref{eqn:numberOfSamples} and $\mathbf v_k^{*}$ be a feasible, optimal solution for~\eqref{eqn:SampleBasedMPCOptimization}. If $\mathbf v_k^{*}$ is applied to the discrete time dynamical system~\eqref{eqn:xUncertainSystem} %
  for a finite horizon of length $T$, then, with at least confidence $1-\delta$, the original chance constraints~\eqref{seqn:probConstraints} are satisfied for $l=0,\dots,T-1$.%
\end{cor}
When applying $\mathbf v_k^*$ in a finite horizon control problem, the terminal region constraint and hard input constraints~\eqref{seqn:inputConstraints} are only met with probability $1-\varepsilon$ and confidence $1-\delta$, as well. When the control law is applied in a receding horizon fashion, then the hard constraints are met as long as~\eqref{eqn:SampleBasedMPCOptimization} remains feasible~\cite{Calafiore2013StochasticMPC}. Note that new scenarios need to be generated at each sampling time.

In~\cite{Schildbach2013_RandomizedSolutions} the results are extended to problems involving multiple chance constraints, which could be applied here to decrease the conservativeness and number of samples. %

\section{Offline Uncertainty Sampling for SMPC} \label{sec:offlineSampling}
In the first part of this section, the difference between online and offline uncertainty sampling in SMPC is illustrated, followed by bounds on the number of samples to guarantee chance constraint satisfaction when the uncertainty is sampled offline.
In the second part, the {off\-line} sampling based SMPC design is derived and the system theoretic properties, recursive feasibility, constraint satisfaction and stability are proven.

\begin{exmp}\label{ex:circConstr}
  Consider the following linear system with parameter $\gamma \in (0,1)$
  \begin{equation*}
    x_{k+1} = 
    \begin{bmatrix}
    1 & \gamma\\ -\gamma & 1
    \end{bmatrix}
    x_k + 
    \begin{bmatrix}
    1 & 0\\ 0 & 1
    \end{bmatrix} u_k
  \end{equation*}
  and a single chance constraint
  \begin{equation}
    \PP\left\{ 
      \begin{bmatrix}
        \cos(\alpha) & \sin(\alpha)
      \end{bmatrix} u_k \le 1
    \right\} \ge 1-\varepsilon
    \label{eqn:exChanceConstr}
  \end{equation}
  where $\alpha \sim \mathcal U[0,2\pi]$.
  For $\varepsilon < 0.5$ the chance constraint set is equal to the constraint $\|u_k\| \le \frac{1}{\cos(\varepsilon\pi)}$. The situation is depicted in Figure~\ref{fig:subsetExample}, where a realization of sampled constraints is given by the blue lines and the area inside the red circle is the feasible region of the chance constraint.

  Let the running cost be given by $\l(x_k,u_k) = \|x_k\|^2 + \|u_k\|^2$, then, for $\|x_k\|$ sufficiently large, the optimal input is at a vertex of the polytope given by the sampled constraints. If we draw new samples at each time step, the probability of the chance constraint being satisfied equals the probability of one \emph{random} vertex being inside the red circle. This holds with confidence $1-\delta$ by Theorem~\ref{thm:scenario}, if the sample complexity is chosen according to~\eqref{eqn:numberOfSamples}. In contrast, when the same set of samples is used and $\|x_0\|$ is sufficiently large, then the optimal solution will switch \emph{deterministically} between the vertices, depending on the system dynamics and cost. For some values of $\gamma$ and the data of the sampled constraint, it will be `trapped' at one vertex.

  Note that the system and constraint can easily be transformed into the form~\eqref{eqn:xUncertainSystem} and~\eqref{seqn:probConstraints}.
\begin{figure}[htb]
  \centering
  \def\circRad{0.10*\textwidth}
  \includegraphics{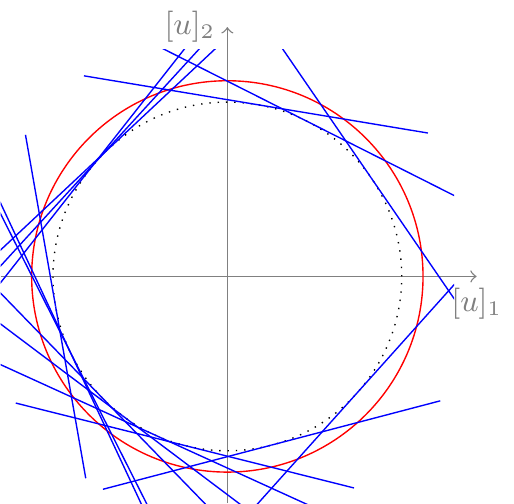}
  \caption{Feasible values $u$ for the chance constraint~\eqref{eqn:exChanceConstr} (solid red), the corresponding robust constraint (dotted gray), and a realization of sampled constraints (solid blue) with one vertex violating the chance constraint.}
  \label{fig:subsetExample}
\end{figure}
\end{exmp}

The previous example highlights the difference between offline and online sampling in Stochastic MPC. %
Let, as defined above, $\mathbb Q_k =  \{\mathbf{q}^{(i)}_k\}_{i=1,\ldots,N_s}$ be the samples used to solve the finite horizon scenario program~\eqref{eqn:SampleBasedMPCOptimization} at time $k$ and define $\mathcal F_k=\{q_0, \mathbb Q_0, \dots, q_k, \mathbb Q_k\}$. 
For $k>0$, the state $x_k$ is a (nonlinear) function not only of all past uncertainty realizations $q_0,\dots,q_{k-1}$, but as well in the sampled scenarios $\mathbb{Q}_0,\dots,\mathbb{Q}_{k-1}$. %
If the scenarios $\mathbb Q_k$ are sampled independently, then for each realization of $\mathcal F_k$, the SMPC optimization~\eqref{eqn:SampleBasedMPCOptimization} reduces to a scenario approximation of a stochastic optimization program with the random variable $q_k$. Using conditional probabilities, the scenario theory and Theorem~\ref{thm:scenario} can be applied, as shown e.g. in~\cite{Calafiore2013StochasticMPC}.
In contrast, if the sampled scenarios are kept constant $\mathbb Q_k = \mathbb Q$, the argument does not hold true and the assumptions of the scenario approach are not satisfied. In fact, due to the system dynamics, each sampled constraint $H_x x_{l|k}^{(i)} \le 1$ depends on \emph{all} sampled scenarios as $x_{l|k}$ is a function in $\mathbb Q$.

Instead of using the scenario results, which guarantee chance constraint satisfaction only for the \emph{optimal solution}, in the following, an approach based on an \emph{inner approximation of the feasible set} is proposed. The number of samples is chosen such that, with confidence $1-\delta$, the sampled constraints are an inner approximation of the chance constraints and hence, with confidence $1-\delta$, each \emph{feasible} solution to the finite horizon scenario program satisfies the original chance constraints.

\subsection{A Sampling Subset Result}\label{ssec:subsetResult}
Let $a$ be a multivariate random variable with realizations in $\R^{1 \times d}$, 
distribution $\PP$ and $\varepsilon,\delta \in (0,1)$. Define the set $\mathbb{C}^{P}$ described by a single chance constraint
\begin{equation}
  \mathbb{C}^{P} = \left\{ x \in \R^d ~|~ \PP\{ a x \le 1 \}\ge1-\varepsilon \right\} \label{eqn:linChanceConstr}
\end{equation}
as the set of those $x$, where the constraint $a x \le 1$ is satisfied with probability $1-\varepsilon$.
Let $a^{(i)}$, $i=1,\dots,N_s$ be $N_s$ iid copies of $a$ and define the second (random) set
\begin{equation}
  \mathbb{C}^{S} = \left\{ x \in \R^d ~|~ a^{(i)}x \le 1, \quad i\in \N_1^{N_s} \right\}. \label{eqn:linCCSampledSet}
\end{equation}
We derive the following proposition on the probability of the set $\mathbb{C}^{S}$%
, described by sampled constraints, being a subset of the set $\mathbb{C}^{P}$%
, described by the chance constraint.
The result is a corollary from~\cite[Corollary 4, Theorem 8]{Alamo2009_RandStrategiesforProbSolOfUncertaintFeasAndOptProbs} and is based on statistical learning theory. The proof on the bound of the VC-Dimension of the class of linear half spaces of the form $\{x ~|~ \theta x\le 1\}$ is given in Appendix~\ref{app:VCDim}.

\begin{prop}\label{prop:subsetProb}
 For any $\varepsilon \in (0,1)$, $\delta \in (0,1)$ and 
 \begin{equation}
   N_s \ge \tilde N(d,\varepsilon,\delta) = \frac{5}{\varepsilon}\left(\ln \frac{4}{\delta} + d \ln \frac{40}{\varepsilon}\right)
   \label{eqn:boundSubsetNumSamples1}
 \end{equation}
 it holds
 \begin{equation*}
   \PP^{N_s}\left\{ \mathbb{C}^S \subseteq \mathbb{C}^P \right\} \ge 1-\delta.
 \end{equation*}
 If $\varepsilon \in (0,0.14)$, the result holds true for the lower bound 
 \begin{equation}
   N_s \ge \frac{4.1}{\varepsilon}\left( \ln\frac{21.64}{\delta} + 4.39 d \log_2\left( \frac{8 \e}{\varepsilon} \right) \right).
   \label{eqn:boundSubsetNumSamples2}
 \end{equation}
\end{prop}
Note that, while Theorem~\ref{thm:scenario} guarantees, with confidence $1-\delta$, satisfaction of the original chance constraint for the optimal solution of the sampled program, Proposition~\ref{prop:subsetProb} implies that, with confidence $1-\delta$, \emph{all} feasible points of the sampled constraints~\eqref{eqn:linCCSampledSet} satisfy the chance constraint in~\eqref{eqn:linChanceConstr}.
In the following section, this result is used to derive a Stochastic MPC scheme based on offline uncertainty sampling which guarantees chance constraint satisfaction and robust recursive feasibility. While the necessary number of samples~\eqref{eqn:boundSubsetNumSamples1} (or \eqref{eqn:boundSubsetNumSamples2}) is larger than~\eqref{eqn:numberOfSamples}, the samples are drawn offline and redundant samples can be removed.

\subsection{Offline Sampling SMPC Design}\label{sec:MainRes}
For the following analysis, we explicitly solve equation \eqref{eqn:xUncertainSystem} with prestabilizing input~\eqref{eqn:prestabContr} for the predicted states $x_{1|k}, \dots, x_{T|k}$ and predicted inputs $u_{0|k}, \dots, u_{T-1|k}$.
With suitable matrices $\Phi^0_{l|k}(\mathbf q_k)$, $\Phi^u_{l|k}(\mathbf q_k)$ %
and $\Gamma_l$ (given in Appendix~\ref{app:matrices}) we get
\begin{equation}
  \begin{aligned}
    x_{l|k}(\mathbf q_k) &= \Phi^0_{l|k}(\mathbf q_k)x_k + \Phi^u_{l|k}(\mathbf q_k) \mathbf v_k\\%
    u_{l|k}(\mathbf q_k) &= K x_{l|k} + v_{l|k} \\
    &= K\Phi^0_{l|k}(\mathbf q_k)x_k + (K\Phi^u_{l|k}(\mathbf q_k)+\Gamma_l) \mathbf v_k.%
  \end{aligned}
  \label{eqn:xuexplicit}
\end{equation}

\paragraph*{Cost Function and Constraint Evaluation}
Given the solution~\eqref{eqn:xuexplicit}, the expected value of the cost~\eqref{eqn:finiteHorizonCostFnc} can be solved explicitly offline, leading to a quadratic, finite horizon cost function 
\begin{equation}
  J_T(x_{k},\mathbf{v}_{k}) = \begin{bmatrix} x_k^\top &\mathbf{v}_k^\top \end{bmatrix} \tilde Q \begin{bmatrix} x_k \\ \mathbf{v}_k \end{bmatrix}%
  \label{eqn:finiteHorizonCostFncDet}
\end{equation}
in the deterministic variables $x_{k}$ and $\mathbf v_k$
with appropriate $\tilde Q$ given in Appendix~\ref{app:matrices}.%

Using the sampling approach developed in Section~\ref{ssec:subsetResult}, an inner approximation for the chance constraints~\eqref{seqn:probConstraints} can be derived in the form of linear constraints on $x_k$ and $\mathbf v_k$.
As before let $\delta \in (0,1)$ be a probabilistic level and for each constraint let $\mathbf q^{(i)}$ be independently drawn samples from $\mathbf Q_k$.
For $l\in\N_1^{T-1}$ and $j\in \N_1^p$, the chance constraints~\eqref{seqn:probConstraints} describe the sets
\begin{equation}
  \X_l^{P,j} = \left\{ x_k, \mathbf v_k ~|~ \PP\{[H_x]_j x_{l|k}(\mathbf q_k) \le 1\} \ge 1-\varepsilon_j \right\}
  \label{eqn:chanceConstrSet}
\end{equation}
for which, with confidence $1-\delta$,
\begin{equation}
  \X_l^{S,j} = \left\{ x_k, \mathbf v_k ~|~ [H_x]_j x_{l|k}(\mathbf q^{(i)}) \le 1, \quad i\in\N_1^{N_{l}} \right\}
  \label{eqn:sampledChanceConstrSet}
\end{equation}
is an inner approximation, if $N_l \ge \tilde N(n+lm,\varepsilon_j,\delta)$.

Similarly, the input and terminal constraints can be approximated. As shown in~\cite{Lorenzen2015_improvedConstrTighteningForSMPC,Lorenzen2015_ConstrTighteningAndStabInSMPC}, for stability and constraint satisfaction it suffices to consider stochastic approximations of the hard constraints on the \emph{predicted} input and terminal constraint. However, the hard input constraint is enforced for the \emph{applied} input $u_{0|k} = v_{0|k}$.
Choose $\varepsilon_h \in (0,1)$ and define
\begin{equation}
  \begin{aligned}
  \U_l^{S,j} &= \left\{ x_k, \mathbf v_k ~|~ [H_u]_j v_{l|k}(\mathbf q^{(i)}) \le 1, \quad i\in\N_1^{N_l^u} \right\}, \\
  \X_T^{S,j} &= \left\{ x_k, \mathbf v_k ~|~ [H_T]_j x_{T|k}(\mathbf q^{(i)}) \le 1, \quad i\in\N_1^{N_T} \right\}
  \end{aligned}
  \label{eqn:sampledInputAndTermConstr}
\end{equation}
for $l\in\N_0^{T-1}$ and $N_T \ge \tilde N(n+Tm,\varepsilon_h,\delta)$,  $N_l^u \ge \tilde N(n+lm,\varepsilon_h,\delta)$.

Note that all sets are described by linear constraints, which, due to the sampling procedure, are in general highly redundant. Since the samples are drawn offline, redundant constraints can be easily removed offline with the following algorithm. %
More sophisticated constraint removal methods already implemented for Matlab can be found e.g. in the MPT toolbox~\cite{MPT3_Toolbox}.

\begin{algorithmic}
  \Require Constraint $H x \le h$
  \Ensure $H \in \R^{n_c \times n}$, $h\in \R^{n_c}$
  \For {$i = 1:n_c$} \newline
    solve
    \begin{equation*}
      \begin{aligned}
        h_i^{*} = \max_x~ & [H]_i x\\
        \text{ s.t. } & [H]_k x \le [h]_k \quad \forall k\in \N_1^{n_c}\setminus i
      \end{aligned}
    \end{equation*}
    \If {$h_i^{*} \leq [h]_i$}
      \State $H \gets H\setminus[H]_i$
      \State $h \gets h\setminus[h]_i$
    \EndIf
  \EndFor \newline
  \Return $H,~h$
\end{algorithmic}

In the following, we assume that the intersection of the sampled constraint sets~\eqref{eqn:sampledChanceConstrSet} and~\eqref{eqn:sampledInputAndTermConstr} with redundant constraints removed is given by
\begin{equation}
  \D = \left\{ x_k, \mathbf v_k ~|~
  \begin{bmatrix} \tilde H & H \end{bmatrix}
  \begin{bmatrix} x_k \\ \mathbf v_k \end{bmatrix} \le h  \right\}.
  \label{eqn:nonredundantConstraints}
\end{equation}

\paragraph*{Recursive Feasibility}
Scenario MPC based on repeatedly solving~\eqref{eqn:SampleBasedMPCOptimization}, cannot guarantee recursive feasibility and hence neither constraint satisfaction nor asymptotic stability of the origin~\cite{Deori2014_CompApproachesToRMPC,Schildbach2014_ScenarioApproachForSMPC}.
In contrast, sampling the uncertainty offline allows to suitably augment the constraints in such a way that recursive feasibility is recovered.

Similar to~\cite{Lorenzen2015_ScenarioBasedStochasticMPC,Lorenzen2015_ConstrTighteningAndStabInSMPC}, a first step constraint is added to~\eqref{eqn:nonredundantConstraints}, as follows.
Let
\begin{equation*}
  \C_T = \left\{ \begin{bmatrix} x_k \\ v_{0|k}  \end{bmatrix} \in \R^{n+m} ~\left|~ 
    \begin{array}[h]{l}
      \exists v_{1|k}, \ldots, v_{T-1|k} \in \R^m, \\
      \text{s.t. } (x_k, \mathbf v_k) \in \D
    \end{array}\right.
\right\}
\end{equation*}
be the $T$-step set and feasible first input. %
The set can be computed via a direct projection, which may introduce numerical difficulties, or recursively with the option for suitable approximations to alleviate the computational complexity in each projection step.
The set $\C_T$ defines the set of feasible states and first inputs of the finite horizon scenario program with given, fixed samples $\mathbf{q}^{(i)}$.
Let %
\begin{equation*}
  \C_{T,x}^\infty = \{x ~|~ H_{\infty}x\le h_{\infty}\}
\end{equation*}
be a (maximal) robust control invariant polytope for the system~\eqref{eqn:xUncertainSystem} with the constraint $(x,u) \in \C_T$. Given the polytopic outer bound $\bar \AB$, this set can be computed via standard recursions; for algorithms and their finite termination see \cite[Section 5.3]{Blanchini2015_SetTheoreticMethodsInControl}\cite{Kerrigan2000_Thesis_RobustConstraintSatisfaction-InvariantSetsAndPredictiveContr} and references therein\footnote{Matlab implementations of those algorithms as part of a toolbox can be found in~\cite{MPT3_Toolbox,Kerrigan2000_Thesis_RobustConstraintSatisfaction-InvariantSetsAndPredictiveContr}.}.

In order for the SMPC optimization to be robustly recursively feasible, the additional constraint set
\begin{equation}
  \D_R = \left\{ x_k, \mathbf v_k ~|~ 
  H_{\infty}A_{cl}^{j} x_k + H_{\infty}B^{j}v_{0|k} \le h_\infty, ~ j\in \N_1^{N_c}
  \right\}
  \label{eqn:firstStepConstraint}
\end{equation}
with $A^j,~B^j$ from Assumption~\ref{ass:uncertaintySetSys} and $A_{cl}^j = A^j + B^jK$ needs to be intersected with~\eqref{eqn:nonredundantConstraints}.

\paragraph*{Sampling-Based SMPC Algorithm}
The complete sampling-based SMPC algorithm can be divided into two parts: (i) an offline computation of the involved sets and removal of redundant constraints and (ii) the repeated online optimization. In the following, we present the algorithm and state its control theoretic properties.%
\\
\emph{Offline:} Compute the expected value~\eqref{eqn:costMatrix} to determine the explicit cost matrix $\tilde Q$ in~\eqref{eqn:finiteHorizonCostFncDet}. Draw a sufficiently large number of samples to determine the sampled constraints~\eqref{eqn:sampledChanceConstrSet} and~\eqref{eqn:sampledInputAndTermConstr}. Remove redundant constraints to get~\eqref{eqn:nonredundantConstraints}. Determine the first step constraint~\eqref{eqn:firstStepConstraint}.
\\
\emph{Online:} For each time step $k \in \N$
\begin{enumerate}
  \item Measure the current state $x_k$.
  \item Determine the minimizer of the quadratic cost~\eqref{eqn:finiteHorizonCostFncDet} subject to the linear constraints~\eqref{eqn:nonredundantConstraints} and~\eqref{eqn:firstStepConstraint} %
    \begin{equation}
      \begin{aligned}
        \mathbf{v}_k^* ~&= \arg \min_{\mathbf{v}_k}
        \begin{bmatrix} x_k^\top &\mathbf{v}_k^\top \end{bmatrix} \tilde Q \begin{bmatrix} x_k \\ \mathbf{v}_k \end{bmatrix}\\
        \text{s.t.} ~
        &(x_k,\mathbf v_k) \in \D \cap \D_R.
      \end{aligned}
      \label{eqn:MPCOptProg}
    \end{equation}
  \item Apply $u_k = Kx_k + v^*_{0|k}$.
\end{enumerate}

\begin{rem}
  The redundant constraints obtained combining~\eqref{eqn:nonredundantConstraints} and~\eqref{eqn:firstStepConstraint} can be removed offline as well. They are kept separate here to emphasize the conceptually different constraints.
\end{rem}

\subsection{Properties of the SMPC Algorithm}\label{sec:MainRes2}
\begin{prop}[Recursive Feasibility] \label{prop:recFeas}
  Let $\V(x_k) = \left\{ \mathbf v_k \in \R^{Tm} ~|~ (x_k,\mathbf v_k) \in \D \cap \D_R \right\}$. If $\mathbf v_k \in \V(x_k)$, then $\V(x_{k+1}) \neq \emptyset$ for every realization $q_k$ and $x_{k+1} = A_{cl}(q_k)x_k + B(q_k)v_{0|k}$.
\end{prop}
\begin{pf}
  From $(x_k, \mathbf v_k) \in \D_R$ it follows $x_{k+1} \in \C_{T,x}^\infty$ robustly and by construction $\C_{T,x}^\infty \subset \{ x ~|~ \V(x)\neq \emptyset\}$.
\end{pf}

 \begin{prop}[Constraint Satisfaction] \label{prop:closedLoopConstraintViolation}
   If $x_0 \in \C_{T,x}^\infty$, the closed-loop system under the proposed SMPC control law satisfies the hard input constraints~\eqref{seqn:inputConstraints} robustly and, with confidence $1-\delta$, the probabilistic constraint~\eqref{seqn:probConstraints} for all $k\ge1$.
 \end{prop}
 \begin{pf}
   Hard input constraint satisfaction follows from robust recursive feasibility (Proposition~\ref{prop:recFeas}) and the constraint $H_u u_{0|k} \le 1$ which does not rely on sampling.

   For all $j=1,\dots,p$ we have $\D \subseteq \X_1^{S,j}$ and by Proposition~\ref{prop:subsetProb}, with confidence $1-\delta$, it holds $\X_1^{S,j} \subseteq \X_1^{P,j}$. Hence, for all feasible $(x_k,\mathbf v_k) \in \D$ the chance constraint $\PP\{ [H_x]_j x_{1|k} \le 1 ~|~ x_k \} \ge 1-\varepsilon$ is satisfied, which suffices for~\eqref{seqn:probConstraints}.
 \end{pf}
 \begin{rem}
   Note that, due to sampling offline, the confidence $1-\delta$ remains the same for all times $k\ge1$. This determines the probability that chance constraint satisfaction does not hold and should therefore be chosen sufficiently small.
 \end{rem}

We have shown that the SMPC algorithm remains recursively feasible, i.e. that a solution to~\eqref{eqn:MPCOptProg} can be found. A stronger assumption usually given, namely that the candidate solution $\tilde v_{l|k+1}=v^*_{l+1|k}$ with $\tilde v_{T-1|k+1}=0$ remains feasible at time $k+1$, is in general impossible to guarantee with sampling-based SMPC. Yet, asymptotic stability of the origin can be proved under a further assumption on the probability that the candidate solution remains feasible.
\begin{assum}\label{ass:upperLowerCostBound}%
  Let $V_T(x)$ be the optimal value function of~\eqref{eqn:MPCOptProg} and 
  let $P_l, P_u \in \R^{n\times n}$, $P_l\succ 0$, $P_u \succ 0$, be such that $x^{\top}P_l x \le V_T(x) \le x^{\top}P_u x$ holds $\forall x \in \C_{T,x}^\infty$.%
\end{assum}%
\noindent
If the candidate solution does not remain feasible, the matrices in Assumption~\ref{ass:upperLowerCostBound} allow to derive bounds on the cost increase.
The matrix $P_l$ is naturally given by the unconstrained infinite horizon cost, while the upper bound can be computed taking into account the vertices of the feasible set $\C_{T,x}^\infty$. %

\begin{prop}[Asymptotic Stability] \label{prop:asStab}
Let $\varepsilon_f$ be an upper bound on the probability that the candidate solution does not remain feasible and let $P_l,~ P_u \in \R^{n\times n}$ satisfy Assumption~\ref{ass:upperLowerCostBound}. If 
  \begin{equation}
    \begin{bmatrix}
      Q - \frac{\varepsilon_f}{1-\varepsilon_f}(A^\top P_u A - P_l) & -\frac{\varepsilon_f}{1-\varepsilon_f}A^\top P_u B\\
      -\frac{\varepsilon_f}{1-\varepsilon_f}B^\top P_u A & R - \frac{\varepsilon_f}{1-\varepsilon_f}(B^\top P_u B)\\
    \end{bmatrix} \succ 0
    \label{eqn:posdefStabCond}
  \end{equation}
holds for all $A,B \in \AB$, the origin is asymptotically stable with probability 1 under the proposed SMPC scheme.
\end{prop}
\begin{rem}
  It suffices to check~\eqref{eqn:posdefStabCond} for the vertices of the set $\bar \AB$, since it can be recast as an LMI in $A$ and $B$ using Schur complement. If $\bar \AB$ is given by interval matrices, the results in~\cite{Alamo2008_NewVertexResultForRobustProbsWithIntervalMatrices} can be applied to further reduce the number of LMIs that need to be checked.
\end{rem}
\begin{pf}
  To prove asymptotic stability, the optimal value function of the online optimization program can be used as a stochastic Lyapunov function. 
  Let $V_T(x_k) = J_T(x_k,\mathbf{v}_{k}^*)$ be the optimal value of~\eqref{eqn:MPCOptProg} at time $k$.

  We first consider the case that the candidate solution remains feasible at time $k+1$.
  Let $\PE\{ V_T(x_{k+1}) | x_k,\allowbreak \mathbf{\tilde v}_{k+1} \text{ feasible}\}$ be the expected optimal value at time $k+1$, conditioning on the state at time $k$ and feasibility of the candidate solution $\tilde v_{l|k+1}=v^*_{l+1|k}$ with $\tilde v_{T-1|k+1}=0$
  \begin{equation*}
    \begin{aligned}
      & \PE\left\{ V_T(x_{k+1})  ~|~ x_k, \mathbf{\tilde v}_{T|k+1}~\text{feasible} \right\} - V_T(x_k) \\
      \le&~ \PE\left\{ J_T(x_{k+1}, \mathbf{\tilde v}_{T|k+1}) ~|~ x_k \right\} - V_T(x_k) \\
      =&~ \PE\left\{\sum_{l=0}^{T-1} \left( \|\tilde x_{l|k+1}\|_Q^2 + \|\tilde u_{l|k+1}\|_R^2  \right) + \|\tilde x_{T|k+1}\|_{P}^2 ~|~ x_k \right\} \\
      & - \PE\left\{ \sum_{l=0}^{T-1} \|x_{l|k}^*\|_Q^2 + \|u_{l|k}^*\|_R^2 + \|x_{T|k}^*\|_P^2 ~|~ x_k \right\} \\
      =&~\PE\left\{ \|x_{T|k}^*\|_{(Q+K^\top R K)}^2 + \|A_{cl}(q_{k+T}) x_{T|k}^*\|_P^2 \right.\\
      &\quad \left. - \|x_{0|k}\|_Q^2 - \|u_{0|k}^*\|_R^2  - \|x_{T|k}^*\|_P^2 ~|~ x_k \right\}  \\
      \le& - \|x_{k}\|_Q^2 - \|u_{k}\|_R^2.
    \end{aligned}
  \end{equation*}
  In case the candidate solution does not remain feasible, Assumption~\ref{ass:upperLowerCostBound} can be employed to compute an upper bound of the Lyapunov function increase
  \begin{equation*}
    \begin{aligned}
      & \PE\left\{ V_T(x_{k+1})  ~|~ x_k, \mathbf{\tilde v}_{T|k+1}~\text{not feasible} \right\} - V_T(x_k) \\
      \le & \max_{(A,B)\in\AB}\|Ax_{k}+Bu_k\|_{P_u}^2 - \|x_{k}\|_{P_l}^2.
    \end{aligned}
  \end{equation*}
  Let $\lambda_{\min}$ be a lower bound of the smallest eigenvalue of the matrix~\eqref{eqn:posdefStabCond} in Proposition~\ref{prop:asStab} for all $(A,B)\in\AB$.
  Then, by the law of total probability, it holds
  \begin{equation*}
    \begin{aligned}
      & \PE\left\{ V_T(x_{k+1})  ~|~ x_k \right\} - V_T(x_k) \\
      \le & -\left(1-\varepsilon_f)( \|x_{k}\|_Q^2 + \|u_{k}\|_R^2 \right) \\
      & \quad
      + \varepsilon_f \left( \max_{(A,B)\in\AB}\|Ax_{k}+Bu_k\|_{P_u}^2 - \|x_{k}\|_{P_l}^2 \right)\\
      \le & -(1-\varepsilon_f)\lambda_{\min}\|x_k\|_2^{2},
    \end{aligned}
  \end{equation*}
  which is a sufficient condition for asymptotic stability with probability 1.%
\end{pf}

\section{Conclusions and Future Work}\label{sec:Concl}
We introduced an offline sampling-based Stochastic MPC scheme for linear systems subject to a parametric disturbance, which can be described by a multivariate random process. 
Unlike previous contributions that are based on online sampling and scenario approximations of the stochastic optimization program, we proposed an easy-to-implement offline sampling scheme for constraint design.
We provided insight into the difference between online and offline sampling in sampling-based SMPC. In particular we have shown that the question of finding a lower bound on the necessary sample complexity can be reduced to the question of a sample approximation being a subset of the original chance constrained set. Results from statistical learning theory have been employed to give explicit bounds, thereby guaranteeing, with a user chosen confidence, constraint satisfaction in closed loop operation.
The approach has the advantage to allow further constraint modifications to guarantee robust recursive feasibility and, under further assumptions, asymptotic stability with probability 1 of the origin.

The main disadvantage compared to Scenario MPC is the increased complexity and the computational requirements in the design phase as well as the restriction to iid disturbance sequences.
Alleviating the latter, e.g. through allowing disturbance sequences that are generated by a dynamical system driven by white noise, and considering unbounded disturbances in combination with probabilistic guarantees of recursive feasibility are future, open research topics.

\appendix
\section{VC-Dimension}\label{app:VCDim}
Note that half-spaces of the form $\{x \in \R^n ~|~ ax \le 1 \}$ always include the origin, which reduces the VC-Dimension from $n+1$ to $n$.
\begin{prop}
  The VC-Dimension of the class of linear half-spaces which include the origin,  $\mathbb H_0 = \{x \in \R^n ~|~ ax \le 1 \}$, $a\in \R^{1 \times n}$ is less or equal than $n$.
\end{prop}
\begin{pf}
  We show that a set $\mathbb X = \{x_1,\dots,x_{n+1}\}$ of cardinality $n+1$ cannot be shattered by $\mathbb H_0$. %
  Let $\mathbb X$ be given. By Radon's Theorem the set $\tilde {\mathbb X} = \{0\} \cup \X$ can be partitioned into two set $\mathbb X_1$ and $\mathbb X_2$ such that $\operatorname{co}({\mathbb X_1}) \cap \operatorname{co}(\mathbb X_2) \neq \emptyset$, where $\operatorname{co}(\X)$ denotes the convex hull of the elements of $\mathbb X$.
  Without loss of generality assume $0 \in \mathbb X_1$. Assume it exists $a \in \R^{1 \times n}$ such that $ax_i \le 1$ for all $x_i \in \mathbb X_1$ and $ax_j > 1$ for all $x_j \in \mathbb X_2$, hence $ax \le 1$ for all $x \in \operatorname{co}(\mathbb X_1)$ and $ax > 1$ for all $x \in \operatorname{co}(\mathbb X_2)$ which contradicts $\operatorname{co}({\mathbb X_1}) \cap \operatorname{co}(\mathbb X_2) \neq \emptyset$.
\end{pf}

\section{Matrices}\label{app:matrices}
For completeness, the matrices used in the paper are given in the following. We use $I_n$ to denote the $n\times n$ identity matrix and $0_{n \times m}$ to denote a zero matrix in $\R^{n \times m}$.
\paragraph*{Solution matrices}
The matrices $\Phi^0_{l|k}$, $\Phi^u_{l|k}$ and $\Gamma_l$ are obtained by solving the dynamics~\eqref{eqn:xUncertainSystem} with prestabilizing input~\eqref{eqn:prestabContr} explicitly for the predicted state $x_{l|k}$ and input $u_{l|k}$.
For $l=0$ we obtain $\Phi^0_{0|k}= I_n$ and $\Phi^u_{0|k}=0_{n\times mT}$.
For $l\ge 1$ with the notation $A_{cl}(q_{l|k}) = A(q_{l|k})+B(q_{l|k})K$
and $A^{cl}_{l|k} = A_{cl}(q_{l|k})$ we have
\begin{equation*}
  \begin{aligned}
  \Phi^{0}_{l|k} &= A^{cl}_{l-1|k} A^{cl}_{l-2|k} \cdots A^{cl}_{0|k}, \\
  \Phi^u_{l|k} &= \begin{bmatrix} A^{cl}_{l-1|k}\cdots A^{cl}_{1|k} B_{0|k} & \dots & B_{l-1|k} & 0_{ln\times (T-l)m} \end{bmatrix}.
  \end{aligned}
\end{equation*}
The matrix $\Gamma_l$ selects the $l-th$ entry in the stack vector $\mathbf v_k$%
\begin{equation*}
  \Gamma_l = \begin{bmatrix}
    0_{m\times lm} & I_m & 0_{m\times (T-l-1)m}
  \end{bmatrix}.
\end{equation*}
\paragraph*{Cost matrix}
Let
\begin{equation*}
  \Phi_T(\mathbf q_k) = 
  \begin{bmatrix}
     \Phi^{0}_{0|k} & \Phi^u_{0|k} \\
     \vdots & \vdots \\
     \Phi^{0}_{T|k} & \Phi^u_{T|k} \\
  \end{bmatrix}, \quad
  \Gamma = 
  \begin{bmatrix}
    0_{mT \times n} & I_{mT \times mT} \\
  \end{bmatrix},
\end{equation*}
$\bar Q = I_T \otimes Q$, $\bar R = I_T \otimes R$ and $\bar K = I_{T} \otimes K$.
The explicit cost matrix $\tilde Q$ in~\eqref{eqn:finiteHorizonCostFncDet} is then given by
\begin{equation}
  \begin{aligned}
  \tilde Q = 
  &\PE\Bigl\{ \Phi_T(\mathbf q_k)^{\top} 
    \begin{bmatrix}
      \bar Q & 0_{nT \times n}\\ 0_{n \times nT} & P
    \end{bmatrix}
    \Phi_T(\mathbf q_k) \\
    & \qquad  +
    \begin{bmatrix}
      \bar K \Phi_{T-1}(\mathbf q_k) + \Gamma
    \end{bmatrix}^{\top}
    \bar R
    \begin{bmatrix}
      \bar K \Phi_{T-1}(\mathbf q_k) + \Gamma
    \end{bmatrix}
  \Bigr\}
  \end{aligned}
  \label{eqn:costMatrix}
\end{equation}
where the expected value can be solved to the desired accuracy using an appropriate numerical integration rule.

\bibliographystyle{plain}
\bibliography{IEEEabrv,GlobalBib}
\end{document}